\documentclass[12pt,preprint]{aastex}

\newcommand{\src}{Swift~J1626.6$-$5156~}

\begin{document} 
\vspace{0.8 in} 

\title{The Orbital Period of \src} 

\author{Altan Baykal\altaffilmark{1}
Ersin {G\"o\u{g}\"u\c{s}}\altaffilmark{2},
S{\i}tk{\i}  \c{C}a\u{g}da\c{s} \.{I}nam\altaffilmark{3},
Tomaso Belloni\altaffilmark{4}
}

\altaffiltext{1}{METU, Physics Department, Ankara 06531 Turkey}
\altaffiltext{2}{Sabanc\i~University, Faculty of Engineering and Natural Sciences, 
Orhanl\i-Tuzla, Istanbul 34956 Turkey}
\altaffiltext{3}{Ba\c{s}kent University, Department of Electrical and
Electronics Engineering, Ankara 06530 Turkey}
\altaffiltext{4}{INAF$-$Osservatorio Astronomico di Brera, Via E. Bianchi 46, 
I-23807 Merate (LC), Italy} 

\email{altan@astroa.physics.metu.edu.tr} 

\begin{abstract} 

We present the discovery of the orbital period of \src. Since its discovery in
2005, the source has been monitored with Rossi X$-$ray Timing Explorer, especially
during the early stage of the outburst and into the X$-$ray modulating episode. 
Using a data span of $\sim$700 days, we obtain the orbital period of the system 
as 132.9 days. We find that the orbit is close to a circular shape with an 
eccentricity 0.08, that is one of the smallest among Be/X$-$ray binary systems. 
Moreover, we find that the timescale of the X$-$ray modulations varied, which led 
to earlier suggestions of orbital periods at about a third and half of the 
orbital period of \src.

\end{abstract}


\section{Introduction} 

Be/X$-$ray systems consist of a neutron star in an eccentric orbit with a 
massive Be star, that is a spectral B type star that displays strong Balmer 
emission lines in its optical spectrum (Coe 2000). Be stars are also observed 
to emit strong infrared radiation. The Balmer emission lines and the infrared 
radiation are usually attributed to the emission from a circumstellar disk 
of material around the Be star. The rotation rates of Be stars are generally high,
sometimes reaching up to $\sim$70\% of the break-up speed. The high rotation 
rates of Be star are thought to take part in forming a circumstellar 
disk (Porter \& Rivinius 2003). In Be/X$-$ray binaries, there is a positive
correlation between the pulse period of the accreting neutron star and the 
orbital period of the binary system (Corbet 1986, Waters\& van Kerkwijk 1989). 

Be/X$-$ray binaries often exhibit X$-$ray outburst episodes. These outbursts are
broadly classified as: {\it normal outbursts} with luminosities around 
10$^{35}$$-$10$^{37}$ erg s$^{-1}$, lasting for a few days to weeks and 
{\it giant outbursts} with luminosities $\gtrsim$ 10$^{38}$ erg s$^{-1}$ and 
lasting for weeks to months (Stella 1986, Negueruela 1998). Normal outbursts
are generally coincident with the time of the periastron passage of the neutron 
star.

\src~was first detected on 2005 December 18 with the Swift Burst Alert
Telescope (BAT) and identified as a transient pulsar with $\sim$15 s pulsations
(Palmer et al 2005). The spin period of the neutron star was refined with
the subsequent Rossi X$-$ray Timing Explorer (RXTE) / Proportional Counter Array
(PCA) observations as  15.37682(5) (Markwardt et al. 2005). At the early phases
of the activation, Reig et al. (2008) observed X$-$ray flare, lasting about
450 s during which the pulsed fraction was seen to increase up to about 70\%.
Following the flare, the average count rate and the pulsed fraction went below 
their pre-flare values for a few hundreds of seconds and indicated a recovery
within a few thousands seconds (Belloni et al. 2006). Optical spectroscopic
observations of the proposed companion (2MASS16263652-5156305, 
USNO-B1.0 0380-0649488) revealed that the star exhibits strong H$\alpha$
emission, indicating that it is a Be star (Negueruela \&  Marco, 2006). 
As the infrared magnitudes of the companion is rather large for a Be star (J=13.5, H=13, K=12.6; Rea et al. 2006), i.e. the star is unusually faint in infrared band, the system is referred to as an
unusual Be/X$-$ray binary system. 
 
Long term monitoring of \src showed strong light curve modulations at timescales
of about 45 days (Reig et al. 2008). This modulation was attributed to variations due to the orbital period of the binary or to a harmonic (Reig et al. 2008). Recently however, DeCesar et al. (2009) reported two periodicities in this system: 47 days and 72.5 days, on approximate 2:3 ratio. 
Based upon their findings, they suggested an orbital period of 23 days. The
issue of the orbital period of \src was not resolved. 
  
In this article, we report the discovery of the orbital period of the Be/X$-$ray
binary system containing \src. In the next section, we describe the
observations. In \S 3, we present our timing analyses and results, and finally
in \S 4, we discuss the implications of our findings.

\section{Observations}

We analyzed data from Proportional Counter Array (PCA) onboard RXTE 
(Jahoda et al 1996) of \src between MJD 53724 and MJD 54410 with a total 
exposure of $\sim 314$ ksec, divided into 289 observations with exposures
between $\sim 1$ ksec and $\sim 2$ ksec. 

The RXTE-PCA is an array of 5 Proportional Counter Units (PCU) operating 
in the 2-60 keV energy range, with a total effective area of approximately 
7000 cm$^2$ and a field of view of $\sim$1$^{\circ}$ FWHM. During the 
observations of \src investigated here, the number of active 
PCUs varied between 1 and 4. 

Between MJD 53724 and MJD 53964, 1-2 ksec long observations were performed 
every 2-3 days. After MJD 53964, observations were sampled in close pairs, 
each of which consists of two consecutive 1-2 ksec long observations separated 
by $\sim0.3-0.6$ days. Each of these observation pairs were apart from each other by $\sim9-10$ days.

\section{Data Analysis and Results}

We generated lightcurves in the 3-20 keV band with a time resolution of 
0.375s using PCA GoodXenon data. Each lightcurve was background 
subtracted using
the background lightcurve generated with the background estimator models 
based upon the rate of very large events, spacecraft activation and 
cosmic X$-$ray emission by mean of the standard PCA analysis tools (pcabackest
in HEADAS). Times of each lightcurve were converted to the barycenter of 
the Solar system. In Figure 3 (top panel), we show the pointing averaged 
net count rate as detected with PCU2. 

In the pulse timing analysis, we have used the harmonic
representation of pulse profiles (Deeter $\&$ Boynton 1985).
In this method, the pulse profiles
were expressed in terms of harmonic series and
cross-correlated with the template pulse profile.
After the 240 days after the main outburst, observations were sampled 
0.3-0.6 days apart from each other. These pair of observations 
were sampled 9-10 days apart from each other. In order to avoid cycle count 
ambiguity due to large gaps in the observation sample, we did not attempt to 
connect cycle counts in phase. For these observations,
we estimated pulse frequencies using the data taken within one day.

\begin{figure}
\begin{center}
\includegraphics[height=5.5in,keepaspectratio=true,angle=-90]{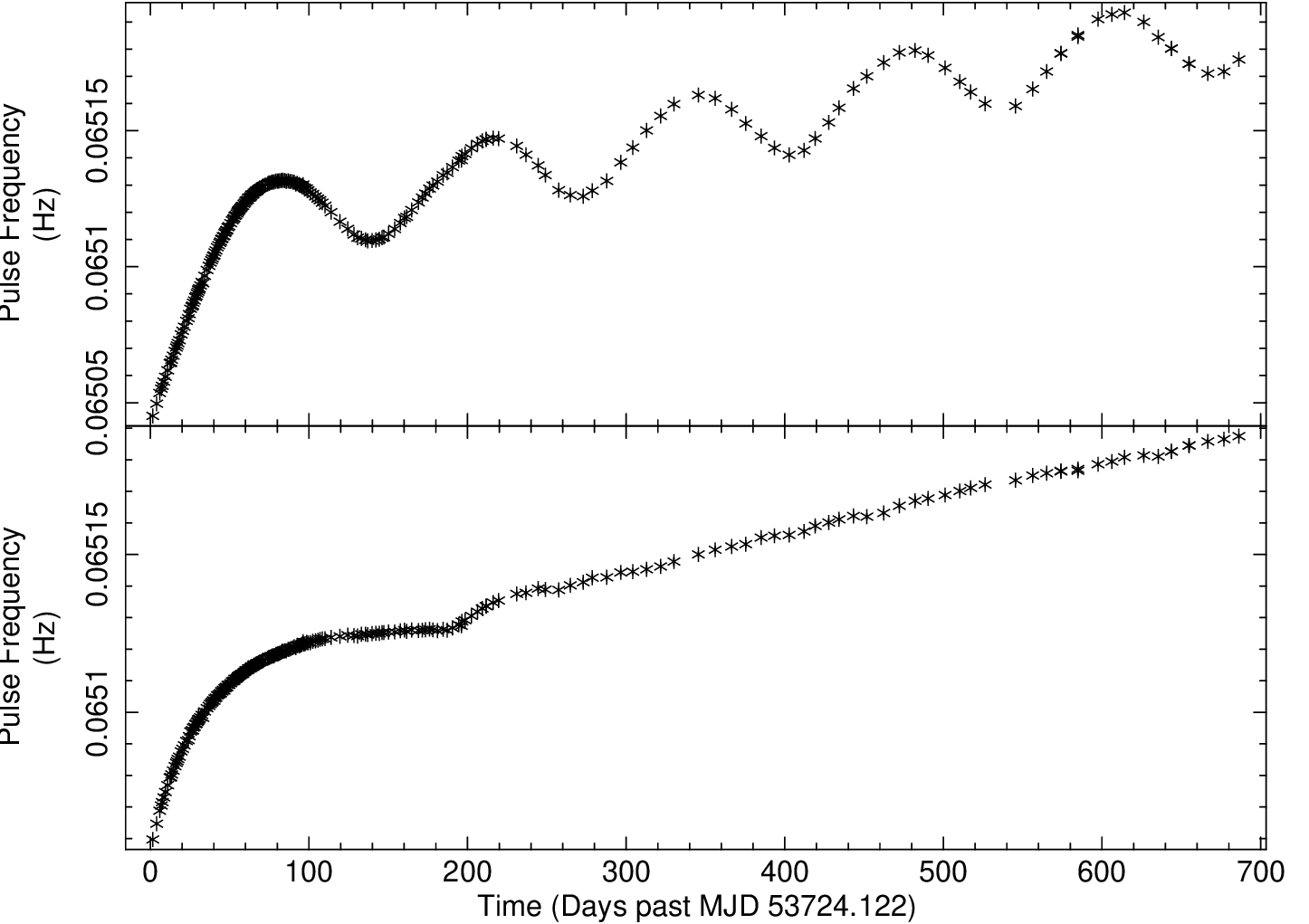}
\end{center}
\small{Figure 1 -- {\bf{(top)}} Pulse frequency evolution of \src. 
{\bf{(bottom)}} The same as above after removing the binary 
orbital motion from the time of arrival of photons.}
\end{figure}

In the upper panel of Figure 1, we present the long term pulse frequency 
evolution of the source. In addition to a spin-up trend, a periodic
modulation due to orbital Doppler effects can be seen.
The Doppler shifted pulse frequency variations, 
spin up rate and its change can be expressed as the first time derivative 
of orbital phase model,

\begin{equation}
\nu = \nu_{0} + \dot \nu (t-t_{0}) + \frac{1}{2} \ddot \nu (t-t_{0})^{2}
+ \frac{2 \pi \nu_{0} }{P_{orbit}} x ( cos (l) + {\bf e}~cos (w) cos (2l) +
                                             {\bf e}~sin (w) sin (2l) ),
\end{equation}
given by Deeter, Boynton, $\&$ Pravdo (1981). Here t$_{0}$ is the mid-time of 
the observation; $\nu_{0}$ is the pulse frequency at t$_{0}$;
$\dot \nu$ and $\ddot \nu$ are the first and second
time derivative of the pulse frequency;
$x=a/c sin(i)$ is the light traveltime
for the projected semimajor axis
(where i is the inclination angle between the line of sight
and the orbital angular momentum vector);
$l=2\pi (t-T_{\pi/2})/P_{orbit}+\pi/2$ is the mean orbital longitude
at t; $T_{\pi/2}$ is the epoch when the mean orbital longitude is equal to
90 $^{\circ}$;
$P_{orbit}$ is the orbital period; e is the eccentricity; and w is the
longitude of periastron. The above expression is fitted to the pulse
pulse frequency time series between 
MJD 53943.74 and MJD 54410.49. In Table 1, we list the parameters of the 
orbital motion, and in Figure 1 (lower panel), we present the frequency evolution
after removal of the orbital effects. 
Figure 2 shows the pulse frequency evolution (after 
removal of the quadratic trend in pulse frequency) together with
the orbit model and their residuals, respectively.
We find that the source was spinning up during the observation with a rate of
$\dot \nu =(1.3062 \pm 0.0017  ) \times 10^{-12}$ Hz s$^{-1}$.
The periodic trend of the pulse frequencies yields
an eccentric orbit (e=$ 0.08 \pm 0.01 $) with an orbital period of
$132.89 \pm 0.03 $ days. 

\begin{figure}
\begin{center}
\includegraphics[width=5.5in,keepaspectratio=true]{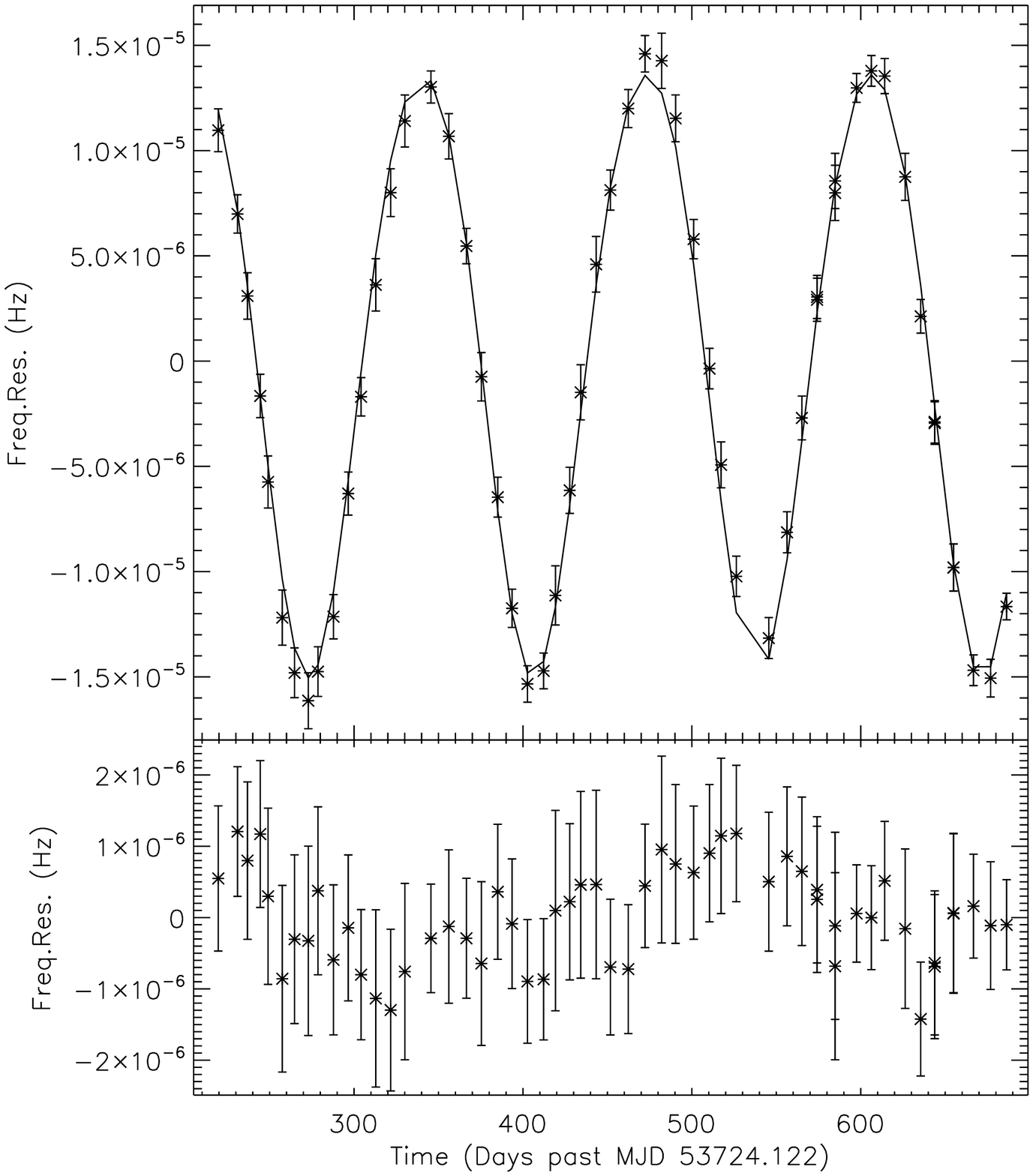}
\end{center}
\small{Figure 2 -- {\bf{(top)}} Pulse frequency time series after removal 
of the quadratic trend in pulse frequency together with the orbit model and 
{\bf{(bottom)}}, their residuals.}
\end{figure}

\begin{table}
\caption{Timing Solution of \src (Valid for MJD 53943.74-54410.49)}
\centering
\begin{tabular}{l|l}\hline\hline
Parameter                   & Value   \\ \hline
Epoch for spin frequency (MJD)  & 54178.24(5)  \\
$\nu$ (Hz)          & 0.065161110(3) \\
$\dot{\nu}$ ($10^{-12}$ Hz.s$^{-1}$)  &  1.31(2) \\
$\ddot{\nu}$ ($10^{-21}$ Hz.s$^{-2}$) &  $-7.5\pm 3.2$ \\
Orbital Period (days)        & 132.89(3) \\
a/c sin i (lt-s)             & 401(5)  \\
Orbital Epoch at $\pi/2$ (MJD)               & 54031.44(5)    \\
Eccentricity                 & 0.08(1)          \\
w (longitude of periastron)  & 340(9)          \\    \hline
\end{tabular}
\label{Table1 -- Parameters}
\end{table}

We also investigated any connection between source intensity and timing
properties of \src. As an intensity indicator, we used count rates in
the 3$-$20 keV range using only PCU2 which had been operating in all observations
used here. Figure 3 shows the count rate history of the source as well as pulse
period evolution after removal of the effects of the orbital motion. We find
that the source exhibits the largest spin-up trend over the interval of the
fastest flux decline in the first $\sim$90 days from MJD 53724 to 83812. When
the source starts to show small scale X$-$ray flaring at around MJD 53814
(the vertical dashed lines in Figure 3), the spin-up trend slows down while the 
underlying X$-$ray intensity continues to
decline. The source exhibits the largest X$-$ray flare starting at $\sim$ MJD
54912 (indicated by the dotted lines in Figure 3). The onset of this flare is 
concident with a major spin-up trend change
from very slow to rather fast. The rapid spin up trend continues throughout
this major X$-$ray flare. After MJD 53965 (marked with the dot-dot-dot-dashed lines
in Figure 3), the spin-up trend remains constant while the source is ongoing 
subsequent X$-$ray flares.

\begin{figure}
\begin{center}
\includegraphics[width=5.5in,keepaspectratio=true]{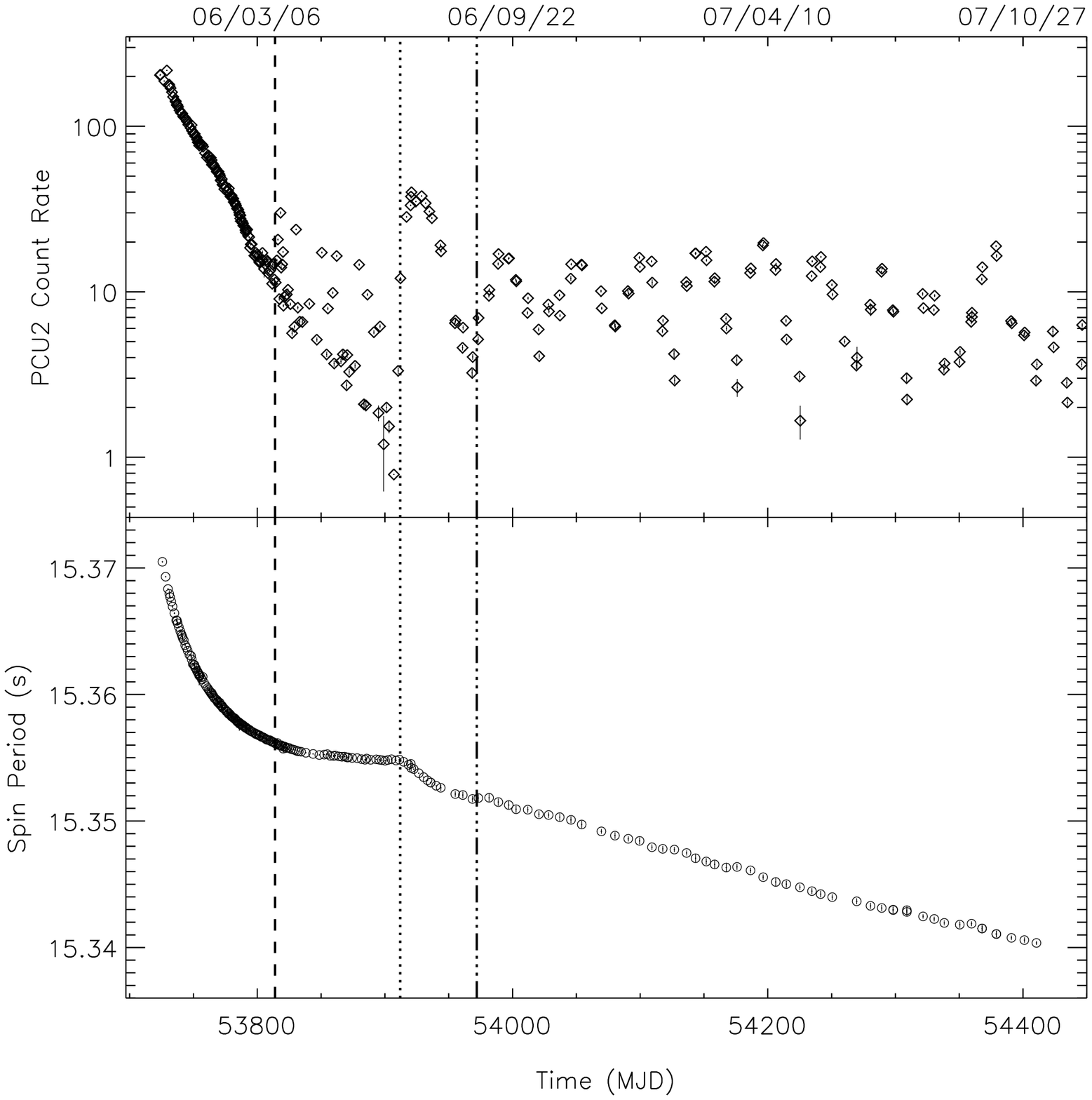}
\end{center}
\small{Figure 3 -- {\bf{(top)}} Count rate history of \src as determined in 
the 3$-$20 keV range using PCU2 data only. {\bf{(bottom)}} Pulse period evolution 
of the source after effects of the orbital motion are removed. Vertical lines are
explained in the text.}
\end{figure}

In order to estimate the timescales of the X$-$ray modulations, we applied a 
Lomb-Scargle algorithm (Lomb 1975, Scargle 1982) for the analysis of 
unevenly-spaced sample to our data.
We used a 200-day long window sliding along the light curve with a time step 
of 7 days. For each stretch, we extracted the period with the highest power. 
The analysis was restricted to the oscillating part of the light curve 
(see bottom panel in Figure 4). The resulting periods, limited to detections 
with a chance probability smaller than 0.1, are plotted in the top panel of 
Figure 4. The size of the points is proportional to minus the logarithm of the 
chance probability, ranging from around $10^{-6}$ around T=400 days (large points) 
to $\sim 10^{-2}$ (small points, T=600 days). The horizontal lines correspond 
to half (dashed) and a third (dotted) of the orbital period of \src. 
We find that the timescale of X$-$ray modulations significantly varies between 
$\sim$45 days and 95 days. The timescale is found to remain nearly constant 
in the interval of 400$-$600 days (see Fugure 4, bottom panel) as it was at 
about 1/3 of the orbital period, and in the interval of 800$-$900 at about half 
the period. There is an abrupt change in timescales of modulations around 
T=610 days as it increased from 45 days ($P_{orb}$/3) to 95 days 
($\sim$$P_{orb}$/3) in about 40 days. The rapid evolution of the timescale 
of modulations can also be seen
in the bottom panel of Figure 4 (notice the separation of the two peaks around 
T=600 and 650 days vs. the separation of the two around 650 and 750 days).

\begin{figure}
\begin{center}
\includegraphics[width=5.5in,keepaspectratio=true]{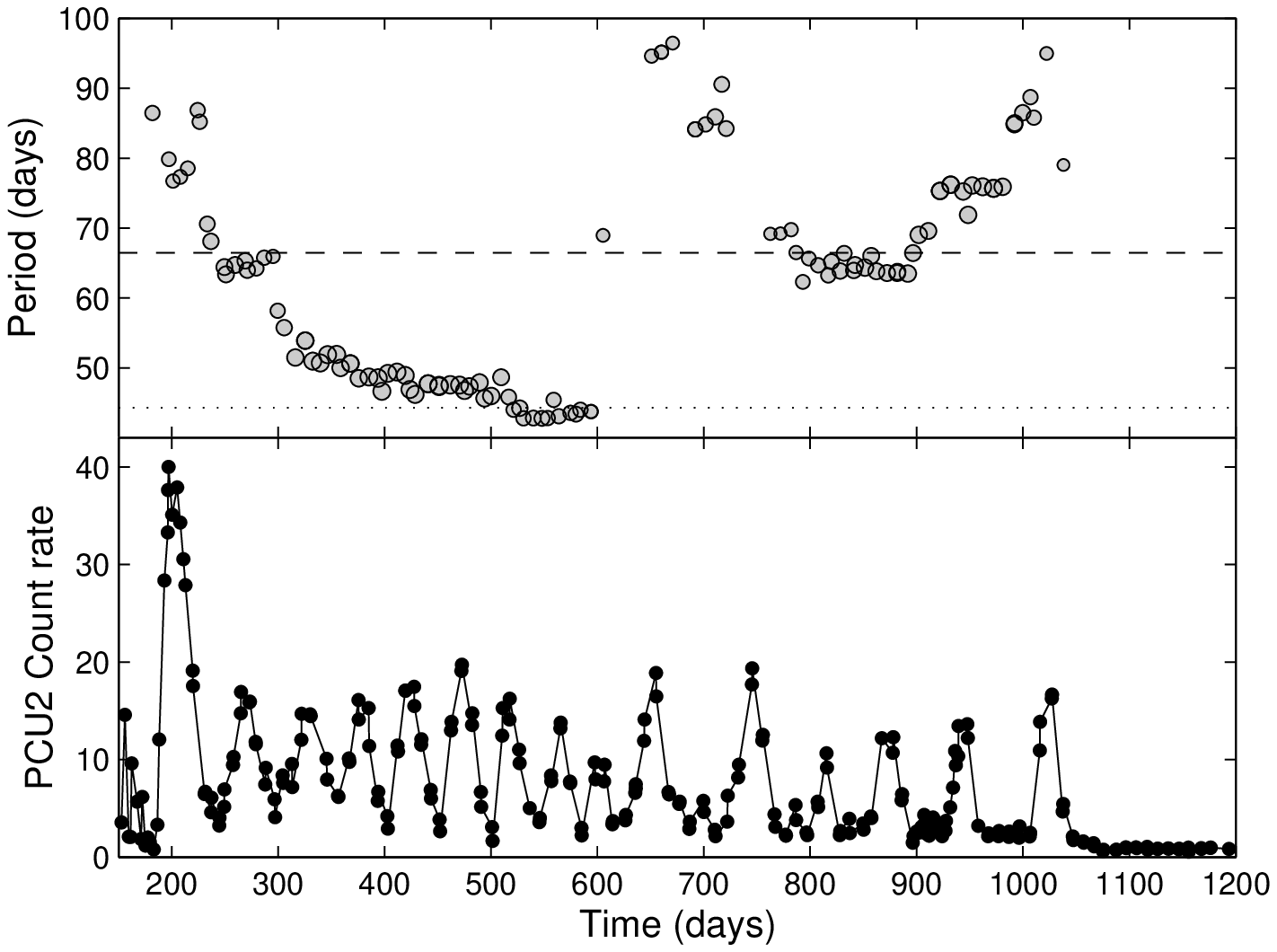}
\end{center}
\small{Figure 4 -- {\bf{(bottom)}} The modulating part of the outburst of \src. 
{\bf{(top)}} The highest-probability period from the sliding Lomb-Scargle analysis 
(see text) as displayed at the middle of the corresponding 200-day long window.
 Symbol size is proportional to minus the logarithm of the chance 
probability. The horizontal lines mark $P_{orb}/2$ (dashed) and $P_{orb}/3$ 
(dotted).}
\end{figure}

\section{Discussion} 

We report the discovery of the orbital period of the binary system harbouring 
\src as 132.9 days. We find that the binary system has an orbital eccentricity 
of 0.08. 

High mass X$-$ray binaries fall into three separate groups when
the pulse periods are compared with the orbital periods (Corbet 1986).
The systems with Be companions show correlations between orbital period
and spin period (Corbet 1986, Waters\& van Kerkwijk 1989), while systems with
giant and supergiant companions fall into two separate regions.
Based on the pulse period and orbital period we discovered, \src falls into the 
region of Be transients in the Corbet diagram.
The derived orbital parameters yield a mass function for the companion,
\begin{equation}
f(M)=\frac{4\pi ^{2}(a_{x} sin i)^{3}}{GP^{2}_{orb}}
=\frac{(M_{c} sin i )^{3}}{(M_{x}+M_{c})^{2}}
\sim 3.9 M_{\odot}.
\end{equation}
For a $\sim$1.4M$_{\odot}$ neutron star mass and orbital inclination angles of
45$^{\circ}$ and 60$^{\circ}$, the mass of the companion star is found to be 
$\sim$13.5$M_{\odot}$ and $\sim$8.3$M_{\odot}$, respectively. In both cases,
the companion star is consistent with being a high mass system.
The optical counterpart of \src shows strong H$\alpha$
emission with an equivalent width of $\sim 40$ $\AA$
(Negueruela $\&$ Marco 2006). This equivalent width implies an
orbital period of the Be/X$-$ray system of the order of $\sim 100-200$ days
which is consistent with the orbital period reported in our work.

"A Be star" is an early type non-supergiant star which has a
circumstellar disk around its equator, possibly formed by fast spin
rotation, non-radial pulsation or magnetic loops (Slettebak 1988).
Be/X$-$ray binary pulsars often show recurrent X$-$ray outbursts. These
X$-$ray outbursts are thought to be mainly due to the
accretion of this circumstellar disk material onto a neutron star
(Negueruela 1998, Inam et al., 2004). The mass accretion increases at
the periastron passages since the density of plasma is greater when
the pulsar is close to the companion star. Therefore it is plausible
to observe an increase in X$-$ray flux during the periastron passage.
This behaviour has been observed in the Be/X$-$ray binaries 4U 0115+63
(Negueruela et al., 1998), KS 1947+300 (Galloway et al., 2004), 2S 1417-624 
(Inam et al. 2004) and EXO 2030+375 (Reig 2008, Baykal et al 2008). In
case of \src, however, we do not find X-ray outbursts coincident with the
periastron passages. This is likely because of the fact that the orbit
is nearly circular (as indicated by very small orbital eccentricity),
therefore, periastron passages do not provide any extra interaction 
between the neutron star and circumstellar disk.  

Although orbital period of \src can also be considered to be typical 
among Be/X$-$ray pulsar binaries, eccentricity of the system is one of 
the smallest among these binaries (Raguzova \& Popov, 2005). The small 
eccentricity of the system can  be as a result of a weak supernova kick 
during formation of the system or a post-supernova circularisation by the 
interaction of the neutron star with the decretion disk of the Be star 
(Martin et al. 2009).  

Our timing analysis of the long-term variability (Figure 4) has shown that 
the situation is more complex than what was reported by DeCesar et al. (2009). 
We find that modulations at timescales of about a half and a third 
of the orbital period are indeed more prominent. For the Be/X-ray pulsar systems with high eccentricity, it is quite natural to observe periodic outbursts near the periastron passages (Okazaki \& Negueruela, 2001). For instance, modulating twice during an orbital revolution has been seen before in 4U~1907+09 (see e.g., in 't Zand, Baykal \& Strohmayer 1998). It could be explained with the  Be star being tilted with respect to the orbital plane. This would cause the neutron star to cross its equatorial disk twice during one orbit, leading to periodic increase in accretion rate. For low eccentric systems such as GS 0834-430 and XTE J1543-568, outbursts are seldom and occur due to mass accumulation until a large perturbation develops (Okazaki \& Negueruela, 2001). The nature of \src is completely different compared to the outbursts of these low eccentric systems. It is more likely that the perturbations in the outer edge of the disk due to the non uniform accretion from the Be companion, propagate to the inner edge of the disk as density waves as suggested in EXO 2030+575 (Wilson et al. 2002, Baykal et al. 2008). These waves lead to small outbursts at the inner edge of the disk. The period of these waves is expected to be related to the resonance size of the accretion disk (Okazaki \& Negueruela, 2001). In eccentric systems, the trigger mechanism of these perturbations is related to the periastron passages, but for \src it should be an intrinsic property of the Be/X-ray pulsar system. There is no definite theory for the trigger mechanism in low eccentric Be/X-ray pulsar systems.

\acknowledgments 

We would like to thank M. Ali Alpar for his valuable comments.
We acknowledge EU FP6 Transfer of Knowledge Project "Astrophysics 
of Neutron Stars" (MTKD-CT-2006-042722). A.B and S.C.I acknowledge research project TBAG 109T748 of the Scientific and
Technological Research Council of Turkey (T\"{U}B\.{I}TAK). TB acknowledges support from ASI grant  ASI I/088/06/0.
\\
\\
{\bf{References}}

{\noindent{Baykal, A., K{\i}z{\i}lo{\u{g}}lu, U., K{\i}z{\i}lo{\u{g}}lu, N., Beklen, E., Ozbey, M., 2008, A\&A, 479, 301}}

{\noindent{Coe, M. J. 2000, in ASP Conf. Ser. 214, Be Phenomenon in Early-Type Stars, ed. M. A. Smith \& H. F. Henrichs (San Francisco: ASP), 656}}

{\noindent{Belloni, T., Homan, J., Campana, S., Markwardt, C. B., Gehrels, N. 
2006, ATel 687}}

{\noindent{Corbet, R.H.D., 1986, MNRAS, 220, 1047}}

{\noindent{DeCesar, M.E., Pottschmidt, K., Wilms, J. 2009, ATel 2036}}

{\noindent{Deeter, J. E., Boynton, P. E., \& Pravdo, S. H. 1981, ApJ, 247, 1003}}

{\noindent{Deeter, J. E., \& Boynton, P. E. 1985, in Proc. Inuyama Workshop on 
Timing Studies of X$-$ray Sources, ed. S. Hayakawa \& F. Nagase (Nagoya: Nagoya 
Univ.), 29}}

{\noindent{Galloway, D.K., Morgan, E.H., Levine, A.M., 2004, ApJ, 613, 1164}}

{\noindent{\.{I}nam S. \c{C}., Baykal A., Scott D. M., Finger M., Swank J., 2004, MNRAS, 349, 173}}

{\noindent{in't Zand J. J. M., Baykal A., Strohmayer T. E., 1998, ApJ, 496, 386}}

{\noindent{Jahoda, K., Swank J. H., Giles A. B., Stark, M. J., Strohmayer, T., 
Zhang, W., \& Morgan, E. H. 1996, Proc. SPIE, 2808, 59}}

{\noindent{Lomb, N.R. 1975, Ap\&SS, 39, 447}}

{\noindent{Markwardt, C. B. \& Swank, J. H. 2005, ATel 679}}

{\noindent{Martin, R.G., Tout, C.A., Pringle, J.E. 2009, MNRAS, 397, 1563}} 

{\noindent{Negueruela, I. 1998, A\&A, 338, 505}}

{\noindent{Negueruela I., Reig, P., Coe, M.J., Fabregat, J., 1998, A\&A, 336, 251}}

{\noindent{Negueruela, I., Marco, A. 2006, ATel 739}}

{\noindent{Okazaki, A., Negueruela, I., 2001, A\&A, 377,161}}

{\noindent{Palmer, D., Barthelmy, S., Cummings, J., et al. 2005, ATel 678}}

{\noindent{Porter, J. M., \& Rivinius, T. 2003, PASP, 115, 1153}}

{\noindent{Raguzova, N.V., Popov, S.B. 2005, A\&AT, 24, 151}}

{\noindent{Rea N., Testa V., Israel G.L., et al. 2006, ATel 713}}

{\noindent{Reig, P., 2008, A\&A, 489, 725}}

{\noindent{Reig, P., Belloni, T., Israel, G.L., Campana, S., Gehrels, N., Homan, J. 
2008, A\&A, 485, 797}}

{\noindent{Scargle, J.D. 1982, ApJ, 263, 835}}

{\noindent{Slettebak A., 1988, PASP 100,770}}

{\noindent{Stella, L., White, N. E., \& Rosner, R. 1986, ApJ, 308, 669}}

{\noindent{Waters, L. B. F. M., \& van Kerkwijk, M. H. 1989, A\&A, 223, 196}}

{\noindent{Wilson, C.A., Finger, M. H., Coe, M.J., Laycock, S., Fabregat, J., 2002, ApJ, 570, 287}}

\end{document}